# SUPPORT POINTS OF LOCALLY OPTIMAL DESIGNS FOR NONLINEAR MODELS WITH TWO PARAMETERS

By Min Yang[1] and John Stufken[2]

*University of Missouri–Columbia and University of Georgia*

We propose a new approach for identifying the support points of a locally optimal design when the model is a nonlinear model. In contrast to the commonly used geometric approach, we use an approach based on algebraic tools. Considerations are restricted to models with two parameters, and the general results are applied to often used special cases, including logistic, probit, double exponential and double reciprocal models for binary data, a loglinear Poisson regression model for count data, and the Michaelis–Menten model. The approach, which is also of value for multi-stage experiments, works both with constrained and unconstrained design regions and is relatively easy to implement.

**1. Introduction.** Generalized linear models (GLMs) and other nonlinear models have found broad applicability during the last decades. Methods of analysis and inference for these models are well established [for GLMs, see, e.g., McCullagh and Nelder (1989), McCulloch and Searle (2001) and Agresti (2002)], but results on optimal designs are more sparse.

In contrast to very general results on optimal designs for linear models with normal errors, many optimality results for nonlinear models are for specific optimality criteria and for specific cases in terms of the model, the design region and the parameters of interest. The significance of our results is their broad applicability, as will be demonstrated in the next sections. As in much of the past work, our method focuses on locally optimal designs. With nonlinear models, information matrices and optimal designs depend on the unknown model parameters, and one way to deal with this is to

Received July 2007.
[1]Supported in part by NSF Grants DMS-07-07013 and DMS-07-48409.
[2]Supported in part by NSF Grant DMS-07-06917.
*AMS 2000 subject classifications.* Primary 62K05; secondary 62J12.
*Key words and phrases.* Design of experiments, optimality, binary response, count data, Poisson model, Michaelis–Menten model, generalized linear model, Loewner order, multi-stage experiment.

identify locally optimal designs based on the best guess of the parameters. There are other ways to address this issue, for example, by using a Bayesian approach [see, e.g., Agin and Chaloner (1999), Chaloner and Larntz (1989) and Chaloner and Verdinelli (1995)]. There is also some interesting work on finding designs that are robust to the best guess of the parameters [see, e.g., Dror and Steinberg (2006)], in which case it would still be useful to know locally optimal designs. As pointed out by Ford, Torsney and Wu (1992), locally optimal designs are important if good initial parameters are available from previous experiments, but can also function as a benchmark for designs chosen to satisfy experimental constraints. Hereafter, we omit the word "locally" for simplicity.

We focus on nonlinear models with two regression parameters. These are the most studied nonlinear models, and our results will unify and generalize many of the available results. We will provide selected relevant references in later sections, and refer the reader to Khuri et al. (2006) for an authoritative recent review on design issues for GLMs.

There can be many design questions that lead to different optimal designs. For example, a design, that is, $D$-optimal for a certain problem may not be $A$-optimal. As another example, a design, that is, optimal for the two parameters may not be optimal for some function of the parameters. Moreover, optimality is a function of the design space, which is in many problems a constrained space. For example, in a toxicity study, a high dose level that exceeds safety limits is not acceptable. All these issues, coupled with the dependence of a locally optimal design on a guess of the parameters, complicate the optimal design problem tremendously. There are consequently very few general results on this topic. Among the exceptions is Biedermann, Dette and Zhu (2006), who obtained a series of excellent results for a specific function of the parameters under a constrained design space and for various models and optimality criteria.

Our strategy will be to identify a class of relatively simple designs so that for any design $d$ that does not belong to this class, there is a design in the class that has an information matrix that dominates that of $d$ in the Loewner ordering. Such a design will then also be no worse than $d$ for most of the common optimality criteria and for many functions of the parameters. The class of designs will depend on the design region—which can be constrained—and the model, and whether a design belongs to the class or not will depend on its support points. With these results, identifying an optimal design for a specific optimality criterion will then either reduce to a simple optimization problem or be solved by using the results of Pukelsheim and Torsney (1991). For further discussion we refer to Biedermann, Dette and Zhu (2006). In Section 6 we will observe that our approach is also useful in the context of multi-stage designs, which is especially helpful in cases where no good initial guess of the parameters is available. We refer to Sitter and Forbes (1997) for further details.

This article is inspired by the work of Mathew and Sinha (2001), who developed a unified approach to tackle optimality problems for the logistic regression model. Our approach successfully characterizes optimal designs under many commonly studied models. Moreover, the results apply for any functions of the original parameters and any commonly used optimality criteria. The results make finding optimal designs for nonlinear models with two parameters a simple task.

For the layout of the remainder of the paper, we will introduce commonly used GLMs and the Michaelis–Menten model in Section 2. In Section 3 we will develop the main tools, which will then be applied to the introduced models in Sections 4 and 5. Section 6 concludes with a brief discussion.

**2. Statistical models and information matrices.** All the models that we will consider have two parameters, which we denote by $\alpha$ and $\beta$, and an explanatory variable $x$. For a given model, the exact optimal design problem consists of selecting distinct values for $x$, say $x_1, \ldots, x_k$, and values for the number of observations $n_i$ at $x_i$ so that the resulting design is best with respect to some optimality criterion for a fixed number of observations $n = \sum_{i=1}^{k} n_i$. The $x_i$'s are the support points of the design. This is a difficult and often intractable optimization problem, which has led to the use of approximate designs in which the $n_i$'s are replaced by $\omega_i$'s that satisfy $\omega_i > 0$ and $\sum_{i=1}^{k} \omega_i = 1$. Thus a design can now be written as $\xi = \{(x_i, \omega_i), i = 1, \ldots, k\}$, and the problem of finding an optimal design becomes, once the support points have been determined, a continuous optimization problem rather than a discrete optimization problem. Identifying support points for an optimal design is therefore extremely important, and the methodology that we develop accomplishes precisely that for optimal designs with a small support size.

For a given design $\xi$ and a two-parameter model for independent observations, the Fisher information matrix for the parameters $(\alpha, \beta)$ can be written as

$$(2.1) \qquad I_\xi(\alpha, \beta) = A^T(\alpha, \beta) C_\xi(\alpha, \beta) A(\alpha, \beta).$$

Here,

$$(2.2) \qquad C_\xi(\alpha, \beta) = \begin{pmatrix} \sum_{i=1}^{k} \omega_i \Psi_1(c_i) & \sum_{i=1}^{k} \omega_i \Psi_2(c_i) \\ \sum_{i=1}^{k} \omega_i \Psi_2(c_i) & \sum_{i=1}^{k} \omega_i \Psi_3(c_i) \end{pmatrix}$$

is a matrix that depends on $(\alpha, \beta)$ (through the $c_i$'s) and on design $\xi$ (through the $\omega_i$'s and $c_i$'s), while $A(\alpha, \beta)$ is a matrix that depends only on $(\alpha, \beta)$. The functions $\Psi_j$ will differ depending on the model. The following examples introduce popular models to illustrate the notation in (2.1) and (2.2).

EXAMPLE 1. In dose-response studies and growth studies, a subject receives a stimulus at a certain level $x$ to study the relationship between the level of the stimulus and a binary response. The dose level in a dose-response study or the level of dilution in a growth study [see McCulloch and Searle (2001), Chapter 5] can be controlled by the experimenter, and a judicious selection of the levels must be made prior to the experiment. With $Y_i$ and $x_i$ as the binary response and the stimulus for the $i$th subject, a basic generalized linear regression model for this situation is of the form

$$(2.3) \qquad \text{Prob}(Y_i = 1) = P(\alpha + \beta x_i).$$

Here, $\alpha$ and $\beta$ are the intercept and slope parameters, and $P(x)$ is a cumulative distribution function. Commonly used models of this form are the logistic, probit, double exponential and double reciprocal models. Most results in the optimal design literature for GLMs are for (2.3).

The information matrix for $(\alpha, \beta)$ for (2.3) is of the form given in (2.1) and (2.2), where we may take $A(\alpha, \beta) = \begin{pmatrix} 1 & -\alpha/\beta \\ 0 & 1/\beta \end{pmatrix}$, $c_i = \alpha + \beta x_i$, $\Psi_1(c_i) = \{P'(c_i)\}^2 / [P(c_i)\{1 - P(c_i)\}]$, $\Psi_2(c_i) = c_i \Psi_1(c_i)$, and $\Psi_3(c_i) = c_i^2 \Psi_1(c_i)$.

EXAMPLE 2. Generalized linear regression models, such as loglinear regression models [Agresti (2002), Chapter 9], can be useful for count data. For example, in a cancer colony-formation assay [Minkin (1993)], the capacity of a drug to reduce the formation of cell colonies is studied. The number of cell colonies observed at a certain concentrate level $x_i$ of the drug is assumed to be a Poisson variable with mean $\theta_i$ and a loglinear model is used to describe the relationship between $\theta_i$ and the concentrate level of the drug $x_i$. Minkin (1993) describes the model as

$$(2.4) \qquad \log \theta_i = \alpha + \beta x_i.$$

Compared to (2.3), the optimal design literature contains fewer results for this model. Nevertheless, the information matrix for $(\alpha, \beta)$ for (2.4) is also of the form given in (2.1) and (2.2) with the same choices as in Example 1, except that now $\Psi_1(c_i) = \exp(c_i)$.

EXAMPLE 3. The Michaelis–Menten model is a nonlinear model that is widely used in the biological sciences. It is given by

$$(2.5) \qquad Y_i = \frac{\alpha x_i}{\beta + x_i} + \varepsilon_i, \qquad \varepsilon_i \sim N(0, \sigma^2),$$

where $\alpha$ and $\beta$ are positive and the explanatory variable $x_i$ can take values in $(0, x_0]$ for some $x_0$. Some results on optimal designs [e.g., Dette and Wong (1999)] are available for this model. The information matrix for $(\alpha, \beta)$ is again of the form given in (2.1) and (2.2), this time with the choices $A(\alpha, \beta) = \begin{pmatrix} 1/\alpha & -1/\beta \\ 0 & 1/\alpha\beta \end{pmatrix}$, $c_i = \alpha x_i/(\beta + x_i)$, $\Psi_1(c_i) = c_i^2$, $\Psi_2(c_i) = c_i^3$ and $\Psi_3(c_i) = c_i^4$.

Our strategy is to identify a class of designs so that for any design $\xi$, and for given $\alpha$ and $\beta$, there is a design $\xi^*$ in the class with $C_{\xi^*}(\alpha, \beta) \geq C_\xi(\alpha, \beta)$. This inequality in the Loewner ordering implies the same inequality for the corresponding information matrices in (2.1), and $\xi^*$ is, for these $(\alpha, \beta)$, locally better than $\xi$ under commonly used optimality criteria, such as $\Phi_p$-optimality, which includes $D$-, $A$- and $E$-optimality. Moreover, if the interest is not in $(\alpha, \beta)$ but in some one-to-one transformation of these parameters, say $(\tau_1, \tau_2)$, then $\xi^*$ is also better than $\xi$ in the Loewner ordering for $(\tau_1, \tau_2)$. This follows easily by observing that the information matrix for $(\tau_1, \tau_2)$, say $J_\xi(\tau_1, \tau_2)$, can be expressed as $J_\xi(\tau_1, \tau_2) = (B^T(\alpha, \beta))^{-1} I_\xi(\alpha, \beta) B^{-1}(\alpha, \beta)$, where $I_\xi(\alpha, \beta)$ is as defined in (2.1) and

$$B(\alpha, \beta) = \begin{pmatrix} \frac{\partial \tau_1}{\partial \alpha} & \frac{\partial \tau_1}{\partial \beta} \\ \frac{\partial \tau_2}{\partial \alpha} & \frac{\partial \tau_2}{\partial \beta} \end{pmatrix}.$$

Observe that this also implies that $\xi^*$ is better than $\xi$ for $\tau_1$ if that is the only parameter function of interest. For this strategy to be useful, the class of designs that we identify must, of course, be a relatively small class consisting of designs that have a small support size.

**3. The main tools.** The main algebraic tools are derived in this section assuming certain properties for the $\Psi_j$'s in (2.2). Applications to specific models will be considered in Sections 4 and 5. We start with the following lemma. In Lemma 1, as well as in Proposition A.1 in the Appendix, $B$ could be $+\infty$, but $A$ and the $c$'s must be finite.

LEMMA 1. *Assume that $\Psi_1(c)$, $\Psi_2(c)$ and $\Psi_3(c)$ are continuous functions on $[A, B]$, that they are three times differentiable on $(A, B]$ and that they satisfy the following conditions on the latter interval:*

(a) $\Psi_1'(c) < 0$;
(b) $(\frac{\Psi_2'(c)}{\Psi_1'(c)})' > 0$;
(c) $((\frac{\Psi_3'(c)}{\Psi_1'(c)})'/(\frac{\Psi_2'(c)}{\Psi_1'(c)})')' > 0$;
(d) $\lim_{c \downarrow A} \frac{\Psi_2'(c)}{\Psi_1'(c)}(\Psi_1(A) - \Psi_1(c)) = 0$.

*Then, for any $c_1$ and $c_2$ with $A < c_1 < c_2 \leq B$ and $0 < \omega < 1$, there exists a unique pair $c_x, \omega_x$, where $c_x \in (c_1, c_2)$ and $0 < \omega_x < 1$, such that*

$$\omega \Psi_1(c_1) + (1-\omega)\Psi_1(c_2) = \omega_x \Psi_1(A) + (1-\omega_x)\Psi_1(c_x), \tag{3.1}$$

$$\omega \Psi_2(c_1) + (1-\omega)\Psi_2(c_2) = \omega_x \Psi_2(A) + (1-\omega_x)\Psi_2(c_x) \tag{3.2}$$

*and*

$$\omega \Psi_3(c_1) + (1-\omega)\Psi_3(c_2) < \omega_x \Psi_3(A) + (1-\omega_x)\Psi_3(c_x). \tag{3.3}$$

*For $\omega = 0$ or $1$, for $c_x \in [c_1, c_2]$, the unique pair that gives equality in (3.1) and (3.2) is $c_x = c_2$ or $c_1$, respectively, and $\omega_x = 0$. This solution also gives equality in (3.3). Furthermore, $c_x$ is a strictly decreasing function of $\omega$.*

PROOF. If $\omega = 0$ or $1$, then it follows from (A.1) and condition (a) that $c_x = c_2$ or $c_1$, respectively, and $\omega_x = 0$. This obviously also gives equality in (3.3).

Next, let $0 < \omega < 1$. We will first show that there is a unique pair $c_x \in (c_1, c_2)$ and $\omega_x \in (0, 1)$ that satisfies (3.1) and (3.2). For $c \in (c_1, c_2)$, define

$$\omega_A(c) = \frac{\omega \Psi_1(c_1) + (1-\omega)\Psi_1(c_2) - \Psi_1(c)}{\Psi_1(A) - \Psi_1(c)}. \tag{3.4}$$

Notice that $\omega_A(c)$ is an increasing function of $c$ by observing that $1 - \omega_A(c)$ is a decreasing function of $c$. For any $c \in (c_1, c_2)$, (3.1) holds for $(c_x, \omega_x) = (c, \omega_A(c))$ [although $\omega_x$ may not be in $(0,1)$]. Define

$$F_1(c) = \omega_A(c)\Psi_2(A) + (1-\omega_A(c))\Psi_2(c) - \omega\Psi_2(c_1) - (1-\omega)\Psi_2(c_2)$$

$$= \frac{\omega \Psi_1(c_1) + (1-\omega)\Psi_1(c_2) - \Psi_1(c)}{\Psi_1(A) - \Psi_1(c)}(\Psi_2(A) - \Psi_2(c))$$

$$+ \Psi_2(c) - \omega\Psi_2(c_1) - (1-\omega)\Psi_2(c_2).$$

For $c = \Psi_1^{-1}(\omega\Psi_1(c_1) + (1-\omega)\Psi_1(c_2))$ [$\Psi_1^{-1}$ exists and $c \in (c_1, c_2)$ since $\Psi_1(c)$ is a monotone function by condition (a)],

$$F_1(c) = \Psi_2(c) - \omega\Psi_2(c_1) - (1-\omega)\Psi_2(c_2) > 0, \tag{3.5}$$

where we have used (A.1). On the other hand, for $c = c_2$, we have

$$F_1(c) = \frac{\omega(\Psi_1(c_1) - \Psi_1(c_2))}{\Psi_1(A) - \Psi_1(c_2)}(\Psi_2(A) - \Psi_2(c_2)) - \omega(\Psi_2(c_1) - \Psi_2(c_2))$$

$$= \omega\left[\frac{\Psi_1(c_1) - \Psi_1(c_2)}{\Psi_1(A) - \Psi_1(c_2)}\Psi_2(A) \right. \tag{3.6}$$

$$\left. + \frac{\Psi_1(A) - \Psi_1(c_1)}{\Psi_1(A) - \Psi_1(c_2)}\Psi_2(c_2) - \Psi_2(c_1)\right] < 0,$$

where we have again used (A.1) in the last step. Since $F_1(c)$ is a continuous function, by (3.5) and (3.6) there must be a $c_x \in (\Psi_1^{-1}(\omega\Psi_1(c_1) + (1-\omega)\Psi_1(c_2)), c_2)$ so that $F_1(c_x) = 0$. Then $c_x$ and $\omega_A(c_x)$, which we will abbreviate to $\omega_x$, satisfy (3.1) and (3.2). Note that $\omega_x \in (0, 1)$ for this choice.

We will now show that the pair $(c_x, \omega_x)$ is unique. Assume that $(c_y, \omega_A(c_y) = \omega_y)$ is another pair that satisfies (3.1) and (3.2). Without loss of generality we may take $c_x < c_y$. By (3.4), this implies $\omega_x < \omega_y$, so that $0 < \frac{\omega_y - \omega_x}{1 - \omega_x} < 1$. Since both $(c_x, \omega_x)$ and $(c_y, \omega_y)$ satisfy (3.1), we have that

$$(3.7) \qquad \Psi_1(c_x) = \frac{\omega_y - \omega_x}{1 - \omega_x}\Psi_1(A) + \frac{1 - \omega_y}{1 - \omega_x}\Psi_1(c_y).$$

By (A.1) this implies

$$(3.8) \qquad \Psi_2(c_x) > \frac{\omega_y - \omega_x}{1 - \omega_x}\Psi_2(A) + \frac{1 - \omega_y}{1 - \omega_x}\Psi_2(c_y).$$

But since both $(c_x, \omega_x)$ and $(c_y, \omega_y)$ satisfy (3.2), we also have

$$(3.9) \qquad \Psi_2(c_x) = \frac{\omega_y - \omega_x}{1 - \omega_x}\Psi_2(A) + \frac{1 - \omega_y}{1 - \omega_x}\Psi_2(c_y),$$

which contradicts (3.8).

The next step consists of showing that the unique pair $(c_x, \omega_x)$ also satisfies inequality (3.3). From (3.1) and (3.2), we obtain

$$
\begin{aligned}
\omega &= ((\Psi_1(c_x) - \Psi_1(c_2))(\Psi_2(c_x) - \Psi_2(A)) - (\Psi_2(c_x) - \Psi_2(c_2)) \\
&\qquad\qquad\qquad\qquad\qquad\qquad\qquad \times (\Psi_1(c_x) - \Psi_1(A))) \\
&\quad \times [(\Psi_1(c_1) - \Psi_1(c_2))(\Psi_2(c_x) - \Psi_2(A)) \\
&\qquad - (\Psi_2(c_1) - \Psi_2(c_2))(\Psi_1(c_x) - \Psi_1(A))]^{-1} \quad \text{and} \\
\omega_x &= ((\Psi_1(c_1) - \Psi_1(c_2))(\Psi_2(c_x) - \Psi_2(c_2)) \\
&\qquad - (\Psi_2(c_1) - \Psi_2(c_2))(\Psi_1(c_x) - \Psi_1(c_2))) \\
&\quad \times [(\Psi_1(c_1) - \Psi_1(c_2))(\Psi_2(c_x) - \Psi_2(A)) \\
&\qquad - (\Psi_2(c_1) - \Psi_2(c_2))(\Psi_1(c_x) - \Psi_1(A))]^{-1}.
\end{aligned}
$$

(3.10)

Note that the common denominator in these expressions is positive by (A.3). Then inequality (3.3) is equivalent to (A.4), and the conclusion follows.

For the final step of the proof we establish that $c_x$ is a strictly decreasing function of $\omega$. For fixed $c_1$ and $c_2$, we have just established that for every $\omega \in (0, 1)$ there exists a unique $c_x$, such that (3.1), (3.2) and (3.3) are satisfied. As seen in (3.10), we can express $\omega$ as a function of $c_x$. This expression also appears in (A.5), and we conclude from (v) of Proposition A.1 that $\omega$ is a strictly decreasing function of $c_x$, which implies also that $c_x$ is a strictly decreasing function of $\omega$. □

The assumptions in Lemma 1 are sufficient to reach the conclusions of the lemma, but certain modifications of the assumptions, which are useful for later applications, can yield the same or similar conclusions. The next two lemmas present variations on Lemma 1. Before we formulate these lemmas we first introduce new terminology. We say that $(\Psi_1(c), \Psi_2(c), \Psi_3(c))$ are *functions of type I* on $[A, B]$ if

(i) $\Psi_1(c)$, $\Psi_2(c)$ and $\Psi_3(c)$ are continuous functions on $[A, B]$ that are three times differentiable on $(A, B]$;

(ii) $\Psi_1'(c)(\frac{\Psi_2'(c)}{\Psi_1'(c)})'((\frac{\Psi_3'(c)}{\Psi_1'(c)})'/(\frac{\Psi_2'(c)}{\Psi_1'(c)})')' < 0$ for $c \in (A, B]$; and

(iii) $\lim_{c \downarrow A} \frac{\Psi_2'(c)}{\Psi_1'(c)}(\Psi_1(A) - \Psi_1(c)) = 0$.

Here $B$ could be $+\infty$, but $A$ must be finite. We say that $(\Psi_1(c), \Psi_2(c), \Psi_3(c))$ are *functions of type II* on $[A, B]$ if

(i) $\Psi_1(c)$, $\Psi_2(c)$ and $\Psi_3(c)$ are continuous functions on $[A, B]$ that are three times differentiable on $[A, B)$;

(ii) $\Psi_1'(c)(\frac{\Psi_2'(c)}{\Psi_1'(c)})'((\frac{\Psi_3'(c)}{\Psi_1'(c)})'/(\frac{\Psi_2'(c)}{\Psi_1'(c)})')' > 0$ for $c \in [A, B)$;

(iii) $\lim_{c \uparrow B} \frac{\Psi_2'(c)}{\Psi_1'(c)}(\Psi_1(B) - \Psi_1(c)) = 0$.

In this case $A$ could be $-\infty$, but $B$ must be finite. The conditions for functions of type I and type II can generally be verified easily; for example, by using symbolic computational software, such as Maple.

Suppose conditions (c) and (d) of Lemma 1 are met, but $\Psi_1'(c) > 0$ and $(\frac{\Psi_2'(c)}{\Psi_1'(c)})' < 0$. It is easily seen that $-\Psi_1(c)$, $\Psi_2(c)$ and $\Psi_3(c)$ satisfy all conditions of the lemma, so that Lemma 1 holds for $-\Psi_1(c)$, $\Psi_2(c)$ and $\Psi_3(c)$. But since (3.1) is an equality, this means that the conclusions of Lemma 1 hold for $\Psi_1(c)$, $\Psi_2(c)$ and $\Psi_3(c)$. Application of Lemma 1 with $\pm\Psi_1(c)$ or $\pm\Psi_2(c)$ yields the following conclusion:

LEMMA 2. *If $(\Psi_1(c), \Psi_2(c), \Psi_3(c))$ are functions of type I on $[A, B]$, then the conclusions of Lemma 1 still hold.*

If $(\Psi_1(c), \Psi_2(c), \Psi_3(c))$ are functions of type II on $[A, B]$, define $\widetilde{\Psi}_1(c) = \Psi_1(-c)$, $\widetilde{\Psi}_2(c) = \Psi_2(-c)$ and $\widetilde{\Psi}_3(c) = \Psi_3(-c)$. It can be verified that $(\widetilde{\Psi}_1(c), \widetilde{\Psi}_2(c), \widetilde{\Psi}_3(c))$ are now functions of type I on $[-B, -A]$. For any $A \leq c < B$, $\Psi_1(c) = \widetilde{\Psi}_1(\widetilde{c})$, where $\widetilde{c} = -c \in (-B, -A]$. Thus, by applying Lemma 2, we have the following result:

LEMMA 3. *Suppose $(\Psi_1(c), \Psi_2(c), \Psi_3(c))$ are functions of type II on $[A, B]$. For any given $A \leq c_1 < c_2 < B$ and $0 < \omega < 1$, there exists a unique pair $c_x, \omega_x$, where $c_x \in (c_1, c_2)$ and $\omega_x \in (0, 1)$, such that (3.1), (3.2) and (3.3) hold with $A$ being replaced by $B$.*

Lemmas 2 and 3 are the basis of this algebraic method. The following two corollaries will be used in our main results. The first one can be derived immediately from Lemma 2.

COROLLARY 1. *Let $(\Psi_1(c), \Psi_2(c), \Psi_3(c))$ be functions of type I on $[A, B]$. For any given $A < c_1 < c_2 \leq B$, $\omega_1 > 0$, and $\omega_2 > 0$, there exists a unique pair $c_x, \omega_x$, where $c_x \in (c_1, c_2)$ and $\omega_x \in (0, \omega_1 + \omega_2)$, such that*

$$\omega_1 \Psi_1(c_1) + \omega_2 \Psi_1(c_2) = \omega_x \Psi_1(A) + (\omega_1 + \omega_2 - \omega_x) \Psi_1(c_x),$$
$$\omega_1 \Psi_2(c_1) + \omega_2 \Psi_2(c_2) = \omega_x \Psi_2(A) + (\omega_1 + \omega_2 - \omega_x) \Psi_2(c_x)$$

*and*

$$\omega_1 \Psi_3(c_1) + \omega_2 \Psi_3(c_2) < \omega_x \Psi_3(A) + (\omega_1 + \omega_2 - \omega_x) \Psi_3(c_x).$$

Applying Corollary 1 repeatedly, we obtain the following result:

COROLLARY 2. *Let $(\Psi_1(c), \Psi_2(c), \Psi_3(c))$ be functions of type I on $[A, B]$. Let $c_i \in (A, B]$ and $\omega_i > 0, i = 1, \ldots, k, k \geq 2$. Then there exists a unique pair $c_x, \omega_x$, where $c_x \in (A, B)$ and $\omega_x \in (0, \sum_{i=1}^{k} \omega_i)$, such that*

$$(3.11) \quad \sum_{i=1}^{k} \omega_i \Psi_1(c_i) = \omega_x \Psi_1(A) + \left(\sum_{i=1}^{k} \omega_i - \omega_x\right) \Psi_1(c_x),$$

$$(3.12) \quad \sum_{i=1}^{k} \omega_i \Psi_2(c_i) = \omega_x \Psi_2(A) + \left(\sum_{i=1}^{k} \omega_i - \omega_x\right) \Psi_2(c_x)$$

*and*

$$(3.13) \quad \sum_{i=1}^{k} \omega_i \Psi_3(c_i) < \omega_x \Psi_3(A) + \left(\sum_{i=1}^{k} \omega_i - \omega_x\right) \Psi_3(c_x).$$

Similarly, Lemma 3 yields the following result:

COROLLARY 3. *Let $(\Psi_1(c), \Psi_2(c), \Psi_3(c))$ be functions of type II on $[A, B]$. Let $c_i \in [A, B)$ and $\omega_i > 0, i = 1, \ldots, k, k \geq 2$. Then there exists a unique pair $c_x, \omega_x$, where $c_x \in (A, B)$ and $\omega_x \in (0, \sum_{i=1}^{k} \omega_i)$, such that*

$$\sum_{i=1}^{k} \omega_i \Psi_1(c_i) = \omega_x \Psi_1(B) + \left(\sum_{i=1}^{k} \omega_i - \omega_x\right) \Psi_1(c_x),$$

$$\sum_{i=1}^{k} \omega_i \Psi_2(c_i) = \omega_x \Psi_2(B) + \left(\sum_{i=1}^{k} \omega_i - \omega_x\right) \Psi_2(c_x)$$

and

$$\sum_{i=1}^{k} \omega_i \Psi_3(c_i) < \omega_x \Psi_3(B) + \left(\sum_{i=1}^{k} \omega_i - \omega_x\right) \Psi_3(c_x).$$

**4. Application to (2.3).** For (2.3) we have that $\Psi_1(c) = \Psi(c)$, $\Psi_2(c) = c\Psi(c)$ and $\Psi_3(c) = c^2\Psi(c)$, where $\Psi(c) = \frac{e^c}{(1+e^c)^2}$ for the logistic model, $\Psi(c) = \frac{\phi^2(c)}{\Phi(c)(1-\Phi(c))}$ for the probit model, $\Psi(c) = \frac{1}{2e^{|c|}-1}$ for the double exponential model and $\Psi(c) = \frac{1}{(1+|c|)^2(2|c|+1)}$ for the double reciprocal model. Here, $\Phi(c)$ and $\phi(c)$ are the c.d.f. and p.d.f. for the standard normal distribution. We observe that all four $\Psi(c)$'s are even and positive functions. By routine algebra, it can be shown that in each case $(\Psi_1(c), \Psi_2(c), \Psi_3(c))$ are type I functions on $[A, B]$ whenever $0 \leq A < B$, including $[0, \infty)$. In addition, for the logistic and probit models, for $c > 0$ it also holds that

$$(4.1) \qquad \left(\frac{\Psi_3'(c)}{\Psi_1'(c)}\right)' > 0.$$

We will use these properties in the derivations of the key results in this section.

We will distinguish between two possible types of constraints on the values of the $c_i$'s. For a symmetric constraint we will assume that $c_i \in [-D, D]$ for some $D > 0$, while $c_i \in [D_1, D_2]$, where $|D_1| \neq |D_2|$, for a nonsymmetric constraint. For the special case that $D = \infty$, there is no constraint at all on the $c_i$'s; if either $D_1 = -\infty$ or $D_2 = \infty$, then there is only a one-sided constraint. We need the following lemma:

LEMMA 4. *Consider (2.3) for the logistic, probit, double exponential or double reciprocal model. Assume $c_i \in [D_1, D_2]$ where $D_1 < 0 < D_2$. For an arbitrary design $\xi = \{(c_i, \omega_i), i = 1, \ldots, k\}$, there exists a design $\tilde{\xi} = \{(c^+, \omega^+), (c^-, \omega^-), (0, 1 - \omega^+ - \omega^-)\}$, such that $C_\xi \leq C_{\tilde{\xi}}$ under the Loewner ordering. Here, $0 < c^+ \leq D_2$ and $D_1 \leq c^- < 0$.*

PROOF. Consider all positive $c_i$'s. If there are none, take $c^+$ as any point with $0 < c^+ \leq D_2$ and $\omega^+ = 0$. If there is one such $c_i$, take this as $c^+$ and take $\omega^+$ to be the corresponding $\omega_i$. For two or more positive $c_i$'s, by Corollary 2 there exist $(c^+, \omega^+)$ and $\omega_{10}$, such that

$$\sum_{c_i > 0} \omega_i \Psi(c_i) = \omega_{10} \Psi(0) + \omega^+ \Psi(c^+),$$

$$\sum_{c_i>0} \omega_i c_i \Psi(c_i) = \omega^+ c^+ \Psi(c^+),$$

$$\sum_{c_i>0} \omega_i c_i^2 \Psi(c_i) < \omega^+ [c^+]^2 \Psi(c^+).$$

Here, $0 < c^+ \leq D_2$ and $\omega_{10} + \omega^+ = \sum_{c_i>0} \omega_i$. Since $\Psi(c)$ is even, a similar result holds for negative $c_i$'s, if any. Let the corresponding values be $(c^-, \omega^-)$, where $D_1 \leq c^- < 0$. Let $\widetilde{\xi} = (c^+, \omega^+), (c^-, \omega^-), (0, 1 - \omega^+ - \omega^-)$. Comparing the two information matrices $C_\xi$ and $C_{\widetilde{\xi}}$, we can see that all elements are the same except that the last diagonal element of the latter exceeds that of the former unless $\xi = \widetilde{\xi}$. The conclusion follows. □

THEOREM 1. *Let $\xi = \{(c_i, \omega_i), i = 1, \ldots, k\}$ with $k \geq 2$ and with $c_i \in [-D, D]$ for some $D > 0$. Then, for the logistic or probit model in (2.3), there is a design $\xi^*$ based on two symmetric points with $C_\xi \leq C_{\xi^*}$. For the double exponential and double reciprocal model, the same conclusion holds except that a third point, namely 0, may have to be included in the support of $\xi^*$.*

PROOF. We first prove the result for the double exponential and double reciprocal models. With $\widetilde{\xi}$ as defined in Lemma 4, it is sufficient to show that there exists a design $\xi^*$ with a support that is based on two symmetric points plus the point 0 and with $C_{\widetilde{\xi}} \leq C_{\xi^*}$. Since $c_i \in [-D, D]$, we have $c^+ \leq D$ and $-c^- \leq D$. If $c^+ = -c^-$, then we take $\xi^* = \widetilde{\xi}$ and the conclusion follows. Otherwise, consider the pair $\{(c^+, \omega^+), (-c^-, \omega^-)\}$ and recall that $\Psi(c^-) = \Psi(-c^-)$. Applying Corollary 1, there exists $(c_x, \omega_x)$, with $c_x$ between $-c^-$ and $c^+$, such that

$$\omega^- \Psi(-c^-) + \omega^+ \Psi(c^+) = (\omega^- + \omega^+ - \omega_x)\Psi(0)$$
$$+ \omega_x \Psi(c_x),$$
(4.2)
$$-\omega^- c^- \Psi(-c^-) + \omega^+ c^+ \Psi(c^+) = \omega_x c_x \Psi(c_x),$$
$$\omega^- [-c^-]^2 \Psi(-c^-) + \omega^+ [c^+]^2 \Psi(c^+) < \omega_x c_x^2 \Psi(c_x).$$

It is clear that $c_x \leq D$. Define

$$2p = \frac{\omega^- c^- \Psi(c^-) + \omega^+ c^+ \Psi(c^+)}{-\omega^- c^- \Psi(-c^-) + \omega^+ c^+ \Psi(c^+)} + 1.$$

Observe that $0 \leq p \leq 1$. Let $\xi^* = \{(c_x, p\omega_x), (-c_x, (1-p)\omega_x), (0, 1 - \omega_x)\}$. Using (4.2) and that $\Psi$ is even, it follows that the corresponding elements of $C_{\widetilde{\xi}}$ and $C_{\xi^*}$ are equal, except that the last diagonal element of $C_{\xi^*}$ could be larger than that of $C_{\widetilde{\xi}}$. Therefore $C_{\widetilde{\xi}} \leq C_{\xi^*}$, and the conclusion follows for the double exponential and double reciprocal models.

These arguments are also valid for the logistic and probit models. Therefore, with $\xi^*$ as the design just constructed, for these models it suffices to show that there exists a design $\xi^0$ based on two symmetric points only such that $C_{\xi^*} \leq C_{\xi^0}$. Since (4.1) holds for the logistic and probit models, by Proposition A.2 there is a unique $c_{x0}$ such that

$$\Psi(c_{x0}) = (1 - \omega_x)\Psi(0) + \omega_x \Psi(c_x),$$

(4.3)

$$[c_{x0}]^2 \Psi(c_{x0}) > \omega_x [c_x]^2 \Psi(c_x).$$

Moreover, from (A.1) it follows that

(4.4) $$c_{x0}\Psi(c_{x0}) > \omega_x c_x \Psi(c_x).$$

Define

$$2\omega_{x0} = \frac{\omega_x(2p-1)c_x\Psi(c_x)}{c_{x0}\Psi(c_{x0})} + 1.$$

Observe that $\omega_{x0} \in (0,1)$. Let $\xi^0 = \{(c_{x0}, \omega_{x0}), (-c_{x0}, 1 - \omega_{x0})\}$. The conclusion follows now as before. □

Almost all results on optimal designs for (2.3) that are currently available in the literature are for the situation that there is no restriction on the design space. For example, for $(\alpha/\beta, \beta)$, optimal designs are found in Abdelbasit and Plackett (1983) and Minkin (1987) (D-optimality); Ford, Torsney and Wu (1992) (c- and D-optimality); Sitter and Wu (1993a, 1993b) (A- and F-optimality) and Sitter and Forbes (1997) (optimal two-stage designs). Mathew and Sinha (2001) provided a unified algebraic approach for the logistic model for deriving A-, D- and E-optimal designs. For $(\alpha, \beta)$, Dette and Haines (1994) investigate E-optimal designs, while Mathew and Sinha (2001) obtained A-optimal designs for the logistic model under the restriction of symmetry, which was removed by Yang (2006). All these results show that optimal designs are based on two symmetric support points for the logistic and probit models, with 0 as a possible additional support point for the double exponential and double reciprocal models. Theorem 1 unifies and extends these results. For example, from Theorem 1 it follows that such results hold as long as the design space is symmetric, for other functions of $\alpha$ and $\beta$, and under more general optimality criteria. We note that Yang (2006) showed that an A-optimal design for $(\alpha, \beta)$ for the double exponential and double reciprocal models could be based on two symmetric points only; this does not contradict Theorem 1, but simply shows that the weight at the point 0 could sometimes be 0 for an optimal design.

The next result shows that for all four models, we can restrict attention to designs with only two support points if the design region is entirely at one side of the origin.

THEOREM 2. *Let $\xi = \{(c_i, \omega_i), i = 1, \ldots, k\}$ with $k \geq 2$ and with $c_i \in [D_1, D_2]$ where either $D_1 \geq 0$ or $D_2 \leq 0$. Then, for the logistic, probit, double exponential or double reciprocal model in (2.3), there is a design $\xi^*$ based on two points with $C_\xi \leq C_{\xi^*}$, and one of the two support points can be taken as $D_1$ if $D_1 \geq 0$ or as $D_2$ if $D_2 \leq 0$.*

We skip the proof since the arguments are similar to those resulting in Lemma 4.

The next result covers the design regions not covered by Theorems 1 and 2.

THEOREM 3. *Let $\xi = \{(c_i, \omega_i), i = 1, \ldots, k\}$ with $k \geq 2$ and with $c_i \in [D_1, D_2]$, where $D_1 < 0 < D_2$ and $-D_1 \neq D_2$. Then for the logistic or probit model in (2.3), there is a design $\xi^*$ based on two support points with $C_\xi \leq C_{\xi^*}$. Moreover, the support points of $\xi^*$ are either two symmetric points; or, if $-D_1 < D_2$, $D_1$ and a point in $(-D_1, D_2]$; or, if $-D_1 > D_2$, $D_2$ and a point in $[D_1, -D_2)$. The same conclusion holds for the double exponential or double reciprocal model, except that 0 may have to be used as a third support point.*

The strategy for a proof of Theorem 3 is not unlike that for Theorem 1, but the details are more onerous and are presented in the Appendix.

While optimal designs for a constrained design space are of great practical value, there are few published papers on this topic. Biedermann, Dette and Zhu (2006) studied $\Phi_p$-optimal designs for $(\sqrt{\lambda}\alpha/\beta, \sqrt{1-\lambda}\beta)$ for $0 < \lambda < 1$ under (2.3). They showed that when the constrained design space is symmetric about 0, then, for the logistic and probit models, the support for an optimal design consists of two symmetric points. For a design space that is not symmetric, depending on the values of $D_1$ and $D_2$, the support points could consist either of two symmetric points or of two points with the one with a smaller absolute value being $D_1$ or $D_2$. Theorem 3 confirms this. Biedermann, Dette and Zhu (2006) further showed that one or both of the support points are end points when the end points are within a certain range. This also could be done using our approach. On the other hand, Biedermann, Dette and Zhu's (2006) approach worked for complementary log-log and skewed logit models, while our approach can handle double exponential and double reciprocal models.

For complementary log-log and skewed logit models, the corresponding $\Psi(c)$ is not an even function. The arguments in Theorems 1 and 3 can therefore not be applied for these two models. However, Corollaries 2 or 3 can be applied for certain design spaces. For example, using $\Psi(c) = \frac{\exp 2c}{\exp(\exp(c))-1}$ for the complementary log-log model, with Maple we find that $(\Psi(c), c\Psi(c),$ $c^2\Psi(c))$ are type II functions for $c \in (-\infty, c_0)$ and type I functions for $c \in (c_0, c_1]$ or $c \in (c_1, \infty)$. Here $c_0$, which is around 0.0491, is the point at which $((\frac{(c^2\Psi(c))'}{\Psi'(c)})'/(\frac{(c\Psi(c))'}{\Psi'(c)})')' = 0$ and $c_1$, which is around 0.4660, is the point at which $\Psi'(c) = 0$. From Corollary 2 or 3, it follows that for any constrained design region within one of the intervals above, all optimal designs are based on two points and one of them is either the lower or upper end-point. (Note that $(\frac{(c\Psi(c))'}{\Psi'(c)})'$ does not exist when $c = c_1$, so that we must separate the two intervals $(c_0, c_1]$ and $(c_1, \infty)$.)

**5. Application to (2.4), (2.5) and other models.** For (2.4) we have that $\Psi_1(c) = e^c$, $\Psi_2(c) = ce^c$ and $\Psi_3(c) = c^2 e^c$. It is easy to show that $(\Psi_1(c), \Psi_2(c), \Psi_3(c))$ are type II functions on $[D_1, D_2]$ for any $D_1 < D_2$. By Corollary 3, we immediately have the following result:

THEOREM 4. *For (2.4), suppose that $\xi$ is a design with support in the design region $[D_1, D_2]$ for some $D_1 < D_2 < \infty$. Then there is a design $\xi^*$ with its support based on two points, one of which is $D_2$, so that $C_\xi \leq C_{\xi^*}$.*

By using the geometric approach, Ford, Torsney and Wu (1992) identified $c$- and $D$-optimal designs under (2.4). They showed that an optimal design has two support points and that one of them is $D_2$. Minkin (1993) studied optimal designs for $1/\beta$ under the same model. In our notation, he assumed $\beta < 0$ and used the design space $(-\infty, \alpha]$. He concluded that the optimal design has two support points, and that one of them is $\alpha$. Theorem 4 confirms and extends these results.

For (2.5) it is easily seen that $\Psi_1(c) = c^2$, $\Psi_2(c) = c^3$ and $\Psi_3(c) = c^4$. It can be shown that $(\Psi_1(c), \Psi_2(c), \Psi_3(c))$ are functions of type II on $[D_1, D_2]$ if $0 < D_1 < D_2$. By Corollary 3, we obtain the following result:

THEOREM 5. *For (2.5), suppose that $\xi$ is a design with support in the design region $[D_1, D_2]$, $0 < D_1 < D_2 < \infty$. Then there is a design $\xi^*$ with its support based on two points, one of which is $D_2$, so that $C_\xi \leq C_{\xi^*}$.*

In our notation, with the design space $(0, \alpha x_0/(\beta + x_0)]$, Dette and Wong (1999) identified $D$- and $E$-optimal designs for $(\alpha, \beta)$ under (2.5) that are two-point designs, with one of them being $\alpha x_0/(\beta + x_0)$. Therefore, Theorem 5 confirms and generalizes these results. Other work on this model can be found in Dette and Biedermann (2003).

Our approach can also be applied to other models. Each time we need to check whether $(\Psi_1(c), \Psi_2(c), \Psi_3(c))$ are functions of type I or type II on an appropriate interval. We will illustrate this for a few examples here. For all of these examples, the information matrix for $(\alpha, \beta)$ can be written as in (2.1) with $\Psi_1(c) = \Psi(c)$, $\Psi_2(c) = c\Psi(c)$ and $\Psi_3(c) = c^2\Psi(c)$ for some function $\Psi(c)$.

Ford, Torsney and Wu (1992) also studied $c$- and $D$-optimal designs for the case $\Psi(c) = c^m$ and the design region $[D_1, D_2]$ for $0 < D_1 < D_2 < \infty$. They considered the cases (i) $m > 0$; (ii) $-2 \leq m \leq 0$; and (iii) $m < -2$. For cases (i) and (ii), except for $m = 0, -1$, and $-2$, it is easily seen that $(\Psi_1(c), \Psi_2(c), \Psi_3(c))$ are functions of type II, while they are of type I for case (iii). Thus, based on our results, we find that optimal designs for cases (i) and (ii), except $m = 0, -1$, and $-2$, can be based on two support points, one of them being $D_2$; for case (iii) the same conclusion holds, except that $D_1$ is now one of the support points. We can further show that, for case (ii), except for $m = 0, -1$, and $-2$, the two support points can be taken as $D_1$ and $D_2$. This can be done by verifying that $\Psi_2'(c)(\frac{\Psi_1'(c)}{\Psi_2'(c)})' > 0$ and $\Psi_2'(c)(\frac{\Psi_3'(c)}{\Psi_2'(c)})' > 0$ and applying (A.16) of Proposition A.3. When $m = 0, -1$, or $-2$, then one of $\Psi_1(c)$, $\Psi_2(c)$, or $\Psi_3(c)$ is constant. The problem becomes simpler in that case, and by applying (A.16) of Proposition A.3 we can show that an optimal design can be based on $D_1$ and $D_2$. Thus this conclusion holds for all $m$ in case (ii). This confirms and extends the conclusions of Ford, Torsney and Wu (1992) for $c$- and $D$-optimal designs (their Tables 2 and 3).

Hedayat, Yan and Pezzuto (2002) identified $c$-optimal designs based on two symmetric points for a nonlinear model with $\Psi(c) = \frac{\exp(2c)}{(1+\exp(c))^4}$. Hedayat, Zhong and Nie (2004) showed that $D$-optimal designs can be based on two symmetric points for a class of two-parameter nonlinear models that includes the following three examples: (i) $\Psi(c) = (1 + c^2)^{-m}$, $m > 1$; (ii) $\Psi(c) = e^{-c^2}$; and (iii) $\Psi(c) = (s + tc^2)^m e^{-lc^2}$, $s, t, l \geq 0$, and $m = 0$ or $m \leq -1$. All of these $\Psi(c)$'s are even functions that satisfy the conditions of Lemma 1 as well as (4.1). Thus they have the same properties as the functions for the logistic and probit models, and the conclusions of Theorems 1, 2 and 3 also hold for all of these models. This confirms and extends the results of Hedayat, Yan and Pezzuto (2002) and Hedayat, Zhong and Nie (2004).

For any nonlinear model, if the corresponding functions $\Psi_1(c)$, $\Psi_2(c)$ and $\Psi_3(c)$ are three times differentiable (which is a condition that is often met), then we may be able to identify intervals so that in each interval $(\Psi_1(c), \Psi_2(c), \Psi_3(c))$ are functions of either type I or type II. If the constrained design space falls entirely within one of the intervals, then either Corollary 2 or 3 can be applied to conclude that an optimal design can be based on two points, with one of them either the lower or upper endpoint of the design region. For example, Hedayat, Yan and Pezzuto (1997) studied $D$-optimal designs for a nonlinear model with $\Psi(c) = \frac{\exp(2rc)}{(1+\exp(c))^{2r+2}}$ for $r > 0$. If, as an example, we take $r = 0.5$, then it can be shown that $(\Psi(c), c\Psi(c), c^2\Psi(c))$ are functions of type II for $c \in (-\infty, c_0)$ and that they are functions of type I when $c \in (c_0, c_1]$ and $c \in (c_1, \infty)$. Here $c_0$, which is around $-0.9131$, is the solution to $((\frac{(c^2\Psi(c))'}{\Psi'(c)})'/(\frac{(c\Psi(c))'}{\Psi'(c)})')' = 0$ and $c_1$, which is around $-0.6931$, is the solution to $\Psi'(c) = 0$.

**6. Discussion.** By Carathéodory's theorem [cf. Silvey (1980)], for a nonlinear model with two parameters, there is an optimal design that is based on at most three support points. However, identifying such points is very challenging. Most studies in this direction are based on the geometric approach, following the seminal work by Elfving (1952), or, especially for $D$-optimal designs, variations on the equivalence theorem by Kiefer and Wolfowitz (1960). Unlike many of these studies, our approach yields very general results that go beyond solving problems on a case by case basis. It helps to identify the support of locally optimal designs for many of the commonly studied models and can be applied for all of the common optimality criteria based on information matrices. It works both with a constrained and unconstrained design region and the conditions needed to reach the conclusions formulated in this paper can be easily verified using symbolic computational software packages.

It is also worthwhile to note that this approach is of value for multi-stage experiments, where an initial experiment may be used to get a better idea about the unknown parameters. At a second or later stage, the question then becomes how to add more design points in an optimal fashion. If $d_1$ denotes the design used so far and $d_2$ the design to be used at the next stage, then the total information matrix is $C_{d_1} + C_{d_2}$. Since the first matrix is fixed, an optimal choice for the second matrix (in the Loewner ordering) is equivalent to making an optimal selection if there had been no prior information through design $d_1$. Therefore, the results in this paper can be used to select $d_2$ by simply ignoring $d_1$.

While the results of this paper are far reaching, we believe that there is potential to extend the approach to nonlinear models with more than two parameters. There are currently very few results on this important practical problem. For example, in a dose-response study, in addition to an explanatory variable $x$, the subjects might be grouped according to race, age, gender, etc. A better model would use these factors to account for heterogeneity in the response between the different groups. We are working on developing tools and results for such problems.

Once the support points for an optimal design have been narrowed down by the methods of this paper, finding an optimal design is a relatively easy problem since we need to consider only the simple structure stated in our results. At worst, a numerical search will now be feasible, but in many cases an analytical solution can be obtained. For example, for the logistic model under (2.3) with a symmetric design region, Mathew and Sinha (2001) conjectured that there is an $A$-optimal design for $(\alpha, \beta)$ that has two symmetric points as its support. Yang (2006) proved this conjecture using a complicated and tedious algebraic approach. However, by our new approach, this result follows immediately and we can easily find such an $A$-optimal design.

For the results with (2.3) and a design region that includes the origin in its interior, we relied on the $\Psi(c)$'s being even functions. We have indicated in Section 4 how partial results can be obtained for the complementary loglog and skewed logit models. Whether our approach can be used to provide complete answers for such models remains an open question.

## APPENDIX

PROPOSITION A.1. *Let $\Psi_1(c)$, $\Psi_2(c)$ and $\Psi_3(c)$ be functions that satisfy the assumptions and conditions formulated in Lemma 1. Then, for fixed $c_1$ and $c_2$ with $A < c_1 < c_2 \leq B$ and any $c \in (A, B]$ and $c_x \in (c_1, c_2)$, the following properties hold:*

(i) *For any $\omega \in (0,1)$, if $\Psi_1(c) = \omega \Psi_1(c_1) + (1-\omega)\Psi_1(c_2)$, then*

(A.1) $$\Psi_2(c) > \omega \Psi_2(c_1) + (1-\omega)\Psi_2(c_2).$$

*This statement remains valid if we allow $c_1 = A$.*

(ii)

(A.2) $$\frac{\Psi_2'(c)}{\Psi_1'(c)} > \frac{\Psi_2(A) - \Psi_2(c)}{\Psi_1(A) - \Psi_1(c)}.$$

(iii)

(A.3) $$(\Psi_1(c_1) - \Psi_1(c_2))(\Psi_2(c_x) - \Psi_2(A))$$
$$- (\Psi_2(c_1) - \Psi_2(c_2))(\Psi_1(c_x) - \Psi_1(A)) > 0.$$

(iv)

(A.4) $$F_2(A, c_1, c_x, c_2)$$
$$:= [\Psi_3(c_1) - \Psi_3(c_2)][(\Psi_1(c_x) - \Psi_1(c_2))(\Psi_2(c_x) - \Psi_2(A))$$
$$- (\Psi_2(c_x) - \Psi_2(c_2))(\Psi_1(c_x) - \Psi_1(A))]$$
$$- [\Psi_3(A) - \Psi_3(c_x)][(\Psi_1(c_1) - \Psi_1(c_2))(\Psi_2(c_x) - \Psi_2(c_2))$$
$$- (\Psi_2(c_1) - \Psi_2(c_2))(\Psi_1(c_x) - \Psi_1(c_2))]$$
$$+ [\Psi_3(c_2) - \Psi_3(c_x)][(\Psi_1(c_1) - \Psi_1(c_2))(\Psi_2(c_x) - \Psi_2(A))$$
$$- (\Psi_2(c_1) - \Psi_2(c_2))(\Psi_1(c_x) - \Psi_1(A))]$$
$$< 0.$$

(v)

$$\begin{aligned}&((\Psi_1(c_x) - \Psi_1(c_2))(\Psi_2(c_x) - \Psi_2(A))\\ &\quad - (\Psi_2(c_x) - \Psi_2(c_2))(\Psi_1(c_x) - \Psi_1(A)))\\ &\quad \times [(\Psi_1(c_1) - \Psi_1(c_2))(\Psi_2(c_x) - \Psi_2(A))\\ &\quad - (\Psi_2(c_1) - \Psi_2(c_2))(\Psi_1(c_x) - \Psi_1(A))]^{-1}\end{aligned}$$ (A.5)

*is a strictly decreasing function of $c_x$.*

PROOF. (i) Fixing $c_1 \in [A, B)$ and $\omega \in (0, 1)$, from condition (a) it follows that for every $c_2 \in (c_1, B]$ there is a unique $c \in (c_1, c_2)$ so that $\Psi_1(c) = \omega \Psi_1(c_1) + (1 - \omega)\Psi_1(c_2)$. Thus, keeping $c_1$ and $\omega$ fixed, we can view $c$ as a function of $c_2$. We have

$$\Psi_1'(c)\frac{dc}{dc_2} = (1-\omega)\Psi_1'(c_2).$$ (A.6)

Define

$$G_1(c_2) = \Psi_2(c) - \omega\Psi_2(c_1) - (1-\omega)\Psi_2(c_2).$$

Using (A.6), we obtain

$$\begin{aligned}G_1'(c_2) &= \Psi_2'(c)\frac{dc}{dc_2} - (1-\omega)\Psi_2'(c_2)\\ &= (1-\omega)\Psi_1'(c_2)\left(\frac{\Psi_2'(c)}{\Psi_1'(c)} - \frac{\Psi_2'(c_2)}{\Psi_1'(c_2)}\right).\end{aligned}$$ (A.7)

Since $c < c_2$, from conditions (a) and (b) we conclude that $G_1'(c_2) > 0$ for $c_2 > c_1$. The result follows by observing that $\lim_{c_2 \downarrow c_1} G_1(c_2) = 0$.

(ii) Define

$$G_2(c) = \frac{\Psi_2'(c)}{\Psi_1'(c)}(\Psi_1(A) - \Psi_1(c)) - (\Psi_2(A) - \Psi_2(c)).$$

Then $G_2'(c) = (\frac{\Psi_2'(c)}{\Psi_1'(c)})'(\Psi_1(A) - \Psi_1(c))$, so that $G_2'(c) > 0$ by (a) and (b). Since $\lim_{c \downarrow A} G_2(c) = 0$ by (d), it follows that $G_2(c) > 0$, which is equivalent to (A.2).

(iii) Define

$$\begin{aligned}G_3(c_1, c_x, c_2) &= (\Psi_1(c_1) - \Psi_1(c_2))(\Psi_2(c_x) - \Psi_2(A))\\ &\quad - (\Psi_2(c_1) - \Psi_2(c_2))(\Psi_1(c_x) - \Psi_1(A)).\end{aligned}$$

For $c_2 > c_x$ we have that

$$\frac{\partial G_3(c_1, c_x, c_2)}{\partial c_2} = \Psi_1'(c_2)(\Psi_1(c_x) - \Psi_1(A))$$
(A.8)
$$\times \left( \frac{\Psi_2'(c_2)}{\Psi_1'(c_2)} - \frac{\Psi_2(A) - \Psi_2(c_x)}{\Psi_1(A) - \Psi_1(c_x)} \right) > 0.$$

The inequality in (A.8) follows from conditions (a) and (b) and (ii) of this proposition. The result follows if we show that $G_4(c_1, c_x) := G_3(c_1, c_x, c_x) > 0$ for $c_x > c_1$. It is easily seen that $G_4(c_1, c_1) = 0$ and

$$\frac{\partial G_4(c_1, c_x)}{\partial c_x} = \Psi_1'(c_x)(\Psi_1(c_1) - \Psi_1(A)) \left( \frac{\Psi_2'(c_x)}{\Psi_1'(c_x)} - \frac{\Psi_2(A) - \Psi_2(c_1)}{\Psi_1(A) - \Psi_1(c_1)} \right).$$

By the same argument as for (A.8), we conclude that $\partial G_4(c_1, c_x)/\partial c_x > 0$, which implies that $G_4(c_1, c_x) > 0$ and yields the desired result.

(iv) Since $F_2(A, c_1, c_x, c_x) = 0$, it suffices to show that $\frac{\partial F_2(A, c_1, c_x, c_2)}{\partial c_2} < 0$. But

$$\frac{\partial F_2(A, c_1, c_x, c_2)}{\partial c_2} = \Psi_3'(c_2)[(\Psi_2(c_1) - \Psi_2(c_x))(\Psi_1(A) - \Psi_1(c_x))$$
$$+ (\Psi_2(c_x) - \Psi_2(A))(\Psi_1(c_1) - \Psi_1(c_x))]$$
$$+ \Psi_2'(c_2)[(\Psi_3(c_1) - \Psi_3(c_x))(\Psi_1(c_x) - \Psi_1(A))$$
$$+ (\Psi_3(c_x) - \Psi_3(A))(\Psi_1(c_x) - \Psi_1(c_1))]$$
$$+ \Psi_1'(c_2)[(\Psi_3(c_1) - \Psi_3(c_x))(\Psi_2(A) - \Psi_2(c_x))$$
$$+ (\Psi_3(c_x) - \Psi_3(A))(\Psi_2(c_1) - \Psi_2(c_x))].$$

This expression is 0 when $c_x = c_1$, so that it suffices to show that $\frac{\partial^2 F_2(A, c_1, c_x, c_2)}{\partial c_2 \partial c_x} < 0$. But

$$\frac{\partial^2 F_2(A, c_1, c_x, c_2)}{\partial c_2 \partial c_x}$$
$$= \Psi_3'(c_2)[\Psi_2'(c_x)(\Psi_1(c_1) - \Psi_1(A)) + \Psi_1'(c_x)(\Psi_2(A) - \Psi_2(c_1))]$$
$$+ \Psi_2'(c_2)[\Psi_3'(c_x)(\Psi_1(A) - \Psi_1(c_1)) + \Psi_1'(c_x)(\Psi_3(c_1) - \Psi_3(A))]$$
$$+ \Psi_1'(c_2)[\Psi_3'(c_x)(\Psi_2(c_1) - \Psi_2(A)) + \Psi_2'(c_x)(\Psi_3(A) - \Psi_3(c_1))].$$

This expression is 0 for $c_1 = A$, so that it suffices to show that $\frac{\partial^3 F_2(A, c_1, c_x, c_2)}{\partial c_2 \partial c_x \partial c_1} < 0$. Simple computation gives

$$\frac{\partial^3 F_2(A, c_1, c_x, c_2)}{\partial c_2 \partial c_x \partial c_1} = \Psi_3'(c_2)[\Psi_2'(c_x)\Psi_1'(c_1) - \Psi_1'(c_x)\Psi_2'(c_1)]$$
$$+ \Psi_2'(c_2)[\Psi_1'(c_x)\Psi_3'(c_1) - \Psi_3'(c_x)\Psi_1'(c_1)]$$

$$+ \Psi'_1(c_2)[\Psi'_3(c_x)\Psi'_2(c_1) - \Psi'_2(c_x)\Psi'_3(c_1)]$$
$$= \Psi'_1(c_1)\Psi'_1(c_x)\Psi'_1(c_2)F_3(c_1, c_x, c_2),$$

where

$$F_3(c_1, c_x, c_2) = \frac{\Psi'_3(c_2)}{\Psi'_1(c_2)}\left(\frac{\Psi'_2(c_x)}{\Psi'_1(c_x)} - \frac{\Psi'_2(c_1)}{\Psi'_1(c_1)}\right) + \frac{\Psi'_2(c_2)}{\Psi'_1(c_2)}\left(\frac{\Psi'_3(c_1)}{\Psi'_1(c_1)} - \frac{\Psi'_3(c_x)}{\Psi'_1(c_x)}\right)$$
$$+ \frac{\Psi'_3(c_x)\Psi'_2(c_1)}{\Psi'_1(c_x)\Psi'_1(c_1)} - \frac{\Psi'_2(c_x)\Psi'_3(c_1)}{\Psi'_1(c_x)\Psi'_1(c_1)}.$$

By condition (a), $\Psi'_1(c_1)\Psi'_1(c_x)\Psi'_1(c_2) < 0$. Thus we need to show that $F_3(c_1, c_x, c_2) > 0$. Since $F_3(c_1, c_x, c_x) = 0$, it suffices to show that $\frac{\partial F_3(c_1,c_x,c_2)}{\partial c_2} > 0$. But

$$\frac{\partial F_3(c_1, c_x, c_2)}{\partial c_2} = \left(\frac{\Psi'_2(c_2)}{\Psi'_1(c_2)}\right)' F_4(c_1, c_x, c_2),$$

where

$$F_4(c_1, c_x, c_2) = \frac{((\Psi'_3(c_2))/(\Psi'_1(c_2)))'}{((\Psi'_2(c_2))/(\Psi'_1(c_2)))'}\left(\frac{\Psi'_2(c_x)}{\Psi'_1(c_x)} - \frac{\Psi'_2(c_1)}{\Psi'_1(c_1)}\right)$$
$$+ \left(\frac{\Psi'_3(c_1)}{\Psi'_1(c_1)} - \frac{\Psi'_3(c_x)}{\Psi'_1(c_x)}\right).$$

Using condition (b), it suffices to show that $F_4(c_1, c_x, c_2) > 0$. But using that $c_1 < c_x < c_2$ and conditions (b) and (c), we obtain that

$$F_4(c_1, c_x, c_2) > \frac{((\Psi'_3(c_x))/(\Psi'_1(c_x)))'}{((\Psi'_2(c_x))/(\Psi'_1(c_x)))'}\left(\frac{\Psi'_2(c_x)}{\Psi'_1(c_x)} - \frac{\Psi'_2(c_1)}{\Psi'_1(c_1)}\right)$$
$$+ \left(\frac{\Psi'_3(c_1)}{\Psi'_1(c_1)} - \frac{\Psi'_3(c_x)}{\Psi'_1(c_x)}\right).$$

The latter expression is 0 when $c_x = c_1$, and by using (b) and (c) we can see that its partial derivative with respect to $c_x$ is positive. This yields the desired conclusion.

(v) The expression in (A.5) is precisely $\omega$ as defined in (3.10). It suffices, therefore, to show that $\partial \omega / \partial c_x < 0$, which is equivalent to showing that $\partial(\frac{\omega}{1-\omega})/\partial c_x < 0$. But $\frac{\omega}{1-\omega} = \frac{G_5(c_1,c_x,c_2)}{G_4(c_1,c_x)}$, where $G_4(c_1, c_x)$ is defined as in the proof of (iii) and

$$G_5(c_1, c_x, c_2) = (\Psi_1(c_x) - \Psi_1(c_2))(\Psi_2(c_x) - \Psi_2(A))$$
$$- (\Psi_2(c_x) - \Psi_2(c_2))(\Psi_1(c_x) - \Psi_1(A)).$$

Clearly $\partial(\frac{G_5(c_1,c_x,c_2)}{G_4(c_1,c_x)})/\partial c_x < 0$ is equivalent to

(A.9) $$\frac{\partial G_5(c_1, c_x, c_2)}{\partial c_x} G_4(c_1, c_x) - G_5(c_1, c_x, c_2)\frac{\partial G_4(c_1, c_x)}{\partial c_x} < 0.$$

Observing that

$$\frac{\partial G_5(c_1, c_x, c_2)}{\partial c_x} = \Psi_2'(c_x)(\Psi_1(A) - \Psi_1(c_2)) - \Psi_1'(c_x)(\Psi_2(A) - \Psi_2(c_2))$$

and using (A.2) and condition (a), we see that $\frac{\partial G_5(c_1, c_x, c_2)}{\partial c_x} < 0$ when $c_2 = c_x$. Since $G_5(c_1, c_x, c_x) = 0$ and $G_4(c_1, c_x) > 0$, we see that (A.9) holds when $c_2 = c_x$. Hence, the result follows if we can show that the left-hand side of (A.9) is a decreasing function of $c_2$. Simple algebra shows that the partial derivative of the left-hand side of (A.9) with respect to $c_2$ is

(A.10)
$$\Psi_1'(c_x)(\Psi_1(c_x) - \Psi_1(A))\left(\frac{\Psi_2'(c_x)}{\Psi_1'(c_x)} - \frac{\Psi_2(A) - \Psi_2(c_x)}{\Psi_1(A) - \Psi_1(c_x)}\right)$$
$$\Psi_1'(c_2)(\Psi_1(A) - \Psi_1(c_1))\left(\frac{\Psi_2'(c_2)}{\Psi_1'(c_2)} - \frac{\Psi_2(A) - \Psi_2(c_1)}{\Psi_1(A) - \Psi_1(c_1)}\right).$$

In (A.10), $\Psi_1'(c_x)$, $\Psi_1(c_x) - \Psi_1(A)$ and $\Psi_1'(c_2)$ are negative while other terms are positive [by (A.2) and conditions (a) and (b)]. Thus (A.10) is negative, which completes the proof. □

PROPOSITION A.2. *Suppose that $\Psi_1(c)$ and $\Psi_3(c)$ are continuous functions on $[A, B]$ and that, for $c \in (A, B)$, they satisfy $\Psi_1'(c) < 0$ and $(\frac{\Psi_3'(c)}{\Psi_1'(c)})' > 0$. Then, for any $A \leq c_1 < c_2 \leq B$ and $\omega \in (0, 1)$, there exists a unique $c \in (c_1, c_2)$ such that*

(A.11)
$$\Psi_1(c) = \omega \Psi_1(c_1) + (1 - \omega)\Psi_1(c_2) \quad and$$
$$\Psi_3(c) > \omega \Psi_3(c_1) + (1 - \omega)\Psi_3(c_2).$$

*Furthermore, c is a strictly decreasing function of $\omega$.*

PROOF. Since $\Psi_1$ is a strictly decreasing function, $\Psi_1^{-1}$ exists. Let $c = \Psi_1^{-1}(\omega \Psi_1(c_1) + (1-\omega)\Psi_1(c_2))$, then clearly $c \in (c_1, c_2)$ and the first equation of (A.11) holds. By the same argument as for (A.1), the inequality in (A.11) holds. The uniqueness of $c$ follows since $\Psi_1$ is strictly decreasing. That $c$ is a strictly decreasing function of $\omega$ is a consequence of $c_1 < c_2$ and the fact that both $\Psi_1$ and $\Psi_1^{-1}$ are strictly decreasing functions. □

We will now present a proof for Theorem 3.

PROOF OF THEOREM 3. Since the $\Psi$'s are even, it suffices to consider the case $-D_1 < D_2$. We will first prove the result for the double exponential and double reciprocal models. It suffices to show that there exists a design as in the statement of the theorem, say $\xi^*$, that satisfies $C_{\widetilde{\xi}} \leq C_{\xi^*}$, where $\widetilde{\xi} =$ $\{(c^+, \omega^+), (c^-, \omega^-), (0, 1-\omega^+-\omega^-)\}$, $0 < c^+ \leq D_2$ is the design in Lemma 4. With $D = \max\{-c^-, c^+\}$, we obtain a design $\xi_0^*$ from Theorem 1 with $C_{\widetilde{\xi}} \leq C_{\xi_0^*}$. However, $\xi_0^*$ may not have its support in $[D_1, D_2]$. If $c_x$ in the proof of Theorem 1 is in $[0, -D_1]$, then we can take $\xi^*$ to be $\xi_0^*$.

Suppose that this is not the case, so that $-c^- < -D_1 < c_x < c^+$. By Corollary 2, the monotonicity of $c_x$ in Lemma 1 and its continuity, there exist $0 < p_0 < 1$ and $0 < \omega_1 < \omega^- + \omega^+ p_0$ such that

(A.12)
$$\omega^- \Psi(-c^-) + \omega^+ p_0 \Psi(c^+) = (\omega^- + \omega^+ p_0 - \omega_1)\Psi(0) + \omega_1 \Psi(-D_1),$$
$$-\omega^- c^- \Psi(-c^-) + \omega^+ p_0 c^+ \Psi(c^+) = -\omega_1 D_1 \Psi(-D_1),$$
$$\omega^- [-c^-]^2 \Psi(-c^-) + \omega^+ p_0 [c^+]^2 \Psi(c^+) < \omega_1 [-D_1]^2 \Psi(-D_1).$$

Define

$$2p_1 = \frac{\omega^- c^- \Psi(c^-) + \omega^+ p_0 c^+ \Psi(c^+)}{-\omega^- c^- \Psi(-c^-) + \omega^+ p_0 c^+ \Psi(c^+)} + 1.$$

Then $0 \leq p_1 \leq 1$, and the design $\xi_1 = \{(D_1, \omega_1(1-p_1)), (-D_1, \omega_1 p_1), (c^+, \omega^+(1-p_0)), (0, 1-\omega^+(1-p_0) - \omega_1)\}$ has a larger information matrix than $\widetilde{\xi}$. By applying Corollary 2, we can further improve the information matrix by replacing the points $-D_1$ and $c^+$ by $0$ and a point $c_x \in (-D_1, c^+)$. The resulting design with support points $D_1$, $c_x$, and $0$ can be taken as design $\xi^*$, giving the conclusion for the double exponential and double reciprocal models.

These arguments are also valid for the logistic and probit models. Thus, for these models, it suffices to show that there exists a design $\xi^0$ based on two symmetric points or on $D_1$ and a point $c_0 \in (-D_1, D_2]$ so that $C_{\xi^*} \leq C_{\xi^0}$, where $\xi^*$ is as in the first part of this proof. If the support of $\xi^*$ consists of the origin and two symmetric points in $[D_1, -D_1]$, then by Theorem 1 we can find a better design based on two symmetric points only. Thus the conclusion follows in that case.

Suppose therefore that $\xi^* = \{(0, \omega_0), (c_x, \omega_x), (D_1, 1 - \omega_0 - \omega_x)\}$, where $c_x > -D_1$. As in the proof of Theorem 1, there exists a $c_{x0} \in (0, c_x)$ such that

$$(\omega_0 + \omega_x)\Psi(c_{x0}) = \omega_0 \Psi(0) + \omega_x \Psi(c_x),$$

and so that two inequalities similar to those in (4.3) and (4.4) hold.

If $c_{x0} \leq -D_1$, then by a similar argument as in the proof of Theorem 1 we can improve design $\xi^*$ by replacing design points $c_x$ and $0$ with design points $c_{x0}$ and $-c_{x0}$ so that the information matrix is larger. The resulting new design is based on $D_1$, $-c_{x0}$, and $c_{x0}$, and by Theorem 1 we can find a design with two symmetric points only that is at least as good.

If, however, $c_{x0} > -D_1$, by the monotonicity property in Proposition A.2 we use that there exists a $p_x \in (0,1)$, such that

$$(\omega_0 + \omega_x p_x)\Psi(-D_1) = \omega_0 \Psi(0) + \omega_x p_x \Psi(c_x).$$

Again by a similar argument as in Theorem 1, we can obtain a design, say $\xi_1$, based on $D_1$, $-D_1$ and $c_x$ that is better than $\xi^*$. Let $\omega_1$ and $\omega_{x0}$ be the weights for $-D_1$ and $c_x$, respectively, in $\xi_1$. For any $q, 0 \leq q \leq 1$, there exists a $c_{x1} \in (-D_1, c_x)$ such that

$$(\omega_1 q + \omega_{x0})\Psi(c_{x1}) = \omega_1 q \Psi(-D_1) + \omega_{x0}\Psi(c_x),$$
(A.13) $\quad(\omega_1 q + \omega_{x0})c_{x1}\Psi(c_{x1}) \geq -\omega_1 q D_1 \Psi(-D_1) + \omega_{x0} c_x \Psi(c_x),$
$$(\omega_1 q + \omega_{x0})c_{x1}^2\Psi(c_{x1}) \geq \omega_1 q [-D_1]^2 \Psi(-D_1) + \omega_{x0} c_x^2 \Psi(c_x).$$

For the two inequalities, equality holds only if $q = 0$. Form a new design, say $\xi_2$, obtained by replacing $(-D_1, \omega_1)$ and $(c_x, \omega_{x0})$ with $(D_1, \omega_1(1-q))$ and $(c_{x1}, \omega_1 q + \omega_{x0})$. Comparing the information matrices for $\xi_1$ and $\xi_2$, by (A.13), for any $q$ the first diagonal elements are the same and $C_{\xi_2}$ has a larger second diagonal element. The difference in the off-diagonal elements of the two matrices is given by

(A.14) $\quad(\omega_1 q + \omega_{x0})c_{x1}\Psi(c_{x1}) + \omega_1(2-q)D_1\Psi(-D_1) - \omega_{x0} c_x \Psi(c_x).$

For $q = 0$, $c_{x1} = c_x$ and the expression in (A.14) is negative. For $q = 1$, by the second inequality in (A.13) the expression in (A.14) is positive. By the first equation of (A.13), $c_{x1}$ is a continuous function of $q$. This implies that (A.14) is also a continuous function of $q$. So there exists a $q \in (0,1)$ such that (A.14) is 0. Let $\xi^0$ be as design $\xi_2$ for that value of $q$. Then $\xi^0$ is better than $\xi^*$ and is based only on $D_1$ and $c_{x1}$. This completes the proof. $\square$

PROPOSITION A.3. *Suppose $\Psi_1(c)$ and $\Psi_3(c)$ are twice differentiable on $[A, B]$ and satisfy*

(A.15) $$\Psi_1'(c)\left(\frac{\Psi_3'(c)}{\Psi_1'(c)}\right)' > 0$$

*for all $c \in [A, B]$. Then, for any $c \in [A, B]$, there exists an $\omega \in (0,1)$, such that*

(A.16)
$$\Psi_1(c) = \omega \Psi_1(A) + (1-\omega)\Psi_1(B) \quad and$$
$$\Psi_3(c) \leq \omega \Psi_3(A) + (1-\omega)\Psi_3(B).$$

PROOF. If (A.15) holds, then either $\Psi_1'(c) > 0$ and $(\frac{\Psi_3'(c)}{\Psi_1'(c)})' > 0$ or $\Psi_1'(c) < 0$ and $(\frac{\Psi_3'(c)}{\Psi_1'(c)})' < 0$. Since the second case can be reduced to the first by defining $\widetilde{\Psi}_1(c) = -\Psi_1(c)$, it suffices to consider only the first case.

Since $\Psi_1'(c) > 0$, $\Psi_1(c)$ is a strictly increasing function on $[A, B]$. Define $\omega = \frac{\Psi_1(B) - \Psi_1(c)}{\Psi_1(B) - \Psi_1(A)}$. Then $0 \leq \omega \leq 1$ for any $c \in [A, B]$, and the equality in (A.16) holds. We will show that the inequality in (A.16) also holds. When $c = A$ this is obvious. So take $c > A$. The inequality in (A.16) is equivalent to $G(A, c, B) \geq 0$, where

$$
\begin{aligned}
G(A, c, B) = {} & (\Psi_3(A) - \Psi_3(B))(\Psi_1(B) - \Psi_1(c)) \\
& + (\Psi_3(B) - \Psi_3(c))(\Psi_1(B) - \Psi_1(A)).
\end{aligned}
\tag{A.17}
$$

Since $G(A, c, B) = 0$ when $B = c$, it suffices to show that $\partial G(A, c, B)/\partial B > 0$. But

$$
\begin{aligned}
\frac{\partial G(A, c, B)}{\partial B} &= \Psi_1'(B)(\Psi_1(c) - \Psi_1(A))\left(\frac{\Psi_3'(B)}{\Psi_1'(B)} - \frac{\Psi_3(c) - \Psi_3(A)}{\Psi_1(c) - \Psi_1(A)}\right) \\
&\geq \Psi_1'(B)(\Psi_1(c) - \Psi_1(A))\left(\frac{\Psi_3'(c)}{\Psi_1'(c)} - \frac{\Psi_3(c) - \Psi_3(A)}{\Psi_1(c) - \Psi_1(A)}\right) > 0,
\end{aligned}
$$

where the last inequality follows by the same argument as for (A.2)  □

Department of Statistics  
University of Missouri–Columbia  
Columbia, Missouri 65211-6100  
USA  
E-mail: yangmi@missouri.edu

Department of Statistics  
University of Georgia  
Athens, Georgia 30602  
USA  
E-mail: jstufken@uga.edu